\documentclass[11pt]{article}
\usepackage{amsmath, amssymb, latexsym}

\newtheorem{thm}{\bf Theorem}[section]

\newtheorem{theorem}{\bf Theorem}[section]
\newtheorem{lemma}[thm]{\bf Lemma}        
\newtheorem{prop}[thm]{\bf Proposition}  
\newtheorem{cor}[thm]{\bf Corollary}
\newtheorem{corollary}[thm]{\bf Corollary}

\newtheorem{definition}[thm]{\bf Definition}

\newtheorem{remark}[thm]{\bf Remark}

\newcommand{\x}{\times}

\newcommand{\Bbox}{
{\unskip\nobreak\hfil\penalty50
\hskip1em\hbox{}\nobreak\hfil{\lower .5pt \hbox{$\Box$}}
\parfillskip=0pt \finalhyphendemerits=0 \par}
}

\newcommand{\eop}{
\ifmmode {\hbox{\Bbox}} \else \Bbox \fi
}

\newcommand{\B}{\mathbb{B}}
\newcommand{\C}{\mathcal{C}}
\newcommand{\FA}{\mathsf{FA}}
\newcommand{\lex}{\leq_{\ell}}
\renewcommand{\o}{\mathbf{o}}
\newcommand{\slex}{<_\ell}
\newcommand{\Q}{\mathbb{Q}}

\renewcommand{\L}{\mathcal{L}}
\renewcommand{\P}{\mathcal{P}}
\newcommand{\R}{\mathcal{R}}
\renewcommand{\Q}{\mathcal{Q}}
\newcommand{\LO}{\textsf{LO}}

\newcommand{\zos}{\{0,1\}^*}
\newcommand{\bfo}{\mathbf{1}}

\begin{document}
\title{A Note on Ordinal DFAs}
\author{S.L. Bloom \\
Department of Computer Science\\
Stevens Institute of Technology\\
Hoboken, NJ, USA\\
\and
YiDi Zhang \\
Department of Mathematics \\
Stevens Institute of Technology\\
Hoboken, NJ, USA}
\date{} 
\maketitle

\begin{abstract}
We prove the following theorem.
Suppose that $M$ is a trim DFA on the Boolean alphabet $0,1$. 
The language $\L(M)$ is 
well-ordered by the lexicographic order $\slex$ iff
whenever the non sink states $q,q.0$ are in the same strong
component, then $q.1$ is a sink.  It is easy to
see that this property is sufficient.
In order to show the necessity,
we analyze the behavior of a $\slex$-descending sequence of words.
This property is used to obtain a polynomial time algorithm
to determine, given a DFA $M$, whether 
$\L(M)$ is well-ordered by the lexicographic order.

Last, we apply an argument in \cite{BE,BEa} to give a proof 
that the least nonregular ordinal is $\omega^\omega $.
\end{abstract} 

\section{Introduction}\label{ }

A regular linear ordering is a component of the initial solution
(in the category $\LO$
of linear orderings, see below) of a finite
system of fixed point equations of the form
\begin{eqnarray*}\label{} 
X_i &=& t_i, \quad i=1,\ldots, n,
\end{eqnarray*} 
where each $t_i$ is a term built from the variables $X_1, \ldots, X_n$ using
the constant symbol \textbf{1}, denoting the one point order, and the binary
function symbol $+$, for ordered sum.  
For example, the initial solution 
\begin{eqnarray*}\label{}
X &=& \bfo + X
\end{eqnarray*} 
is the  nonnegative integers, ordered as usual, and the initial solution of
\begin{eqnarray*}\label{}
X &=& X + \bfo + X
\end{eqnarray*} 
is the rationals, ordered as usual.  (It is known 
that such systems
have initial solutions in $\LO$ \cite{BEmezei,Adamek,Wand}.)
When the ordering is well-founded,
it is a regular well-ordering.

Any countable (regular) linear ordering is
isomorphic to a (regular) subset of words ordered lexicographically.
(See \cite{Courcelle, CourcelleFund}.)  
We consider the question: which trim
deterministic finite automata, or DFAs, $M$ have the property that
the language $\L(M)$, ordered lexicographically, is well-ordered?
(Such DFAs are the ``ordinal DFAs'' of the title.)

As a consequence, we obtain
a polynomial time algorithm to determine,
given a DFA $M$, whether $(\L(M),\slex)$ is well-ordered.

If $\alpha $ is a regular ordinal, there is a trim
DFA $M$  such that $(\L(M),\slex)$ 
has order-type $\alpha $. 
We obtain the known result that $\alpha$ is less than $\omega^\omega $
by adopting to DFAs the technique in \cite{BE} which was applied
to context-free grammars.

\section{Preliminaries}\label{ }

We review some well-known concepts to establish our terminology.
A linearly ordered set $(L,<)$ is a set equipped with a strict
linear ordering, i.e., a transitive, irreflexive relation such
that for $x,y \in L$, exactly one of $x=y,\ x<y,\ y<x$ holds.
Here, we will assume that any linearly ordered set is at most
countable.   A \textbf{morphism}
$  \varphi: (L,<_1) \to (L',<_2)$ of linearly ordered
sets is a function that preserves the ordering:
if $x<_1 y$ then $\varphi(x) <_2 \varphi(y)$, and thus $\varphi $ is injective.
Thus, the linearly ordered sets form a category $\LO$.  Two linearly ordered
sets are isomorphic if they are isomorphic in this category.
A linearly ordered set $(L,<)$ is \textbf{well-ordered} if every nonempty subset
of $L$ has a least element.   The \textbf{order-type} 
$\o(L,<)$ of a linearly ordered 
set is the isomorphism class of $(L,<)$.
A (countable) \textbf{ordinal} is 
the order-type of a well-ordered set. 
 
If $(L,<_1)$ and $(L',<_2)$ are linearly ordered sets, the ordered sum
\begin{eqnarray*}
 (L,<_1) + (L',<_2)
\end{eqnarray*}
is the linearly ordered set obtained by defining all points in $L$ to be
less than all points in $L'$, and otherwise keeping the original orders.
More generally, if for each $n \geq 0$, $(L_n,<_n)$ is  a linearly ordered
set, then the \textbf{ordered sum}
\begin{eqnarray*}
  (L_0, <_0) + (L_1, <_1) + \ldots 
\end{eqnarray*}
is the set $\bigcup_n L_n \x \{n\} $ ordered as follows:
\begin{eqnarray*}
  (x,i) < (y,j) &\iff & i<j \text{ or } i=j \text{ and } x<_i y.
\end{eqnarray*}
If a set $\Sigma $ is linearly ordered, the \textbf{lexicographic order} on
the set of words on $\Sigma $, $\Sigma^*$, is defined 
for $u,v \in \Sigma^*$ by
\begin{eqnarray*}\label{}
u \lex v &\iff & u \leq _ p v \text{ or } u <_s v, 
\end{eqnarray*}
where $\leq_p$ is the \textbf{prefix order} and $<_s$ is the \textbf{strict order}:
\begin{eqnarray*}\label{}
u \leq_p v &\iff & v = w u, \text{ for some } w \in \Sigma^*, \text{ and }  \\
u <_s v &\iff & u = x \sigma_1 w \text{ and } v = x \sigma_2 w', \text{ for some } x,w,w' \in \Sigma^* \text{ and } \\
  &&  \sigma_1 <  \sigma_2 \text{ in } \Sigma .
\end{eqnarray*} 
We write $u \slex v$ if $u \neq v$ and $u \lex v$.  
If $u$ is the word $b_0 b_1 \ldots b_{k-1}$ whose
length $|u|$ is $k$ and if
$0 \leq i \leq j < k$, we write
$
u[i \ldots j]
$
for the subword
$b_i \ldots b_j$ of $u$.  Also, we write $(u)_i$ for the $i$-th letter
$b_i$ of $u$. In particular, $(u)_i = u[i \ldots i]$.  

The next Proposition recalls some elementary facts.
\begin{prop}\label{elem prop}
\begin{enumerate}
\item For any two distinct words $u,v$ with $|u| \leq |v|$,  
either $u \leq_p v$, or $u <_s v$ or $v <_s u$.
\item   If $u \slex v$, then $wu <_s wv$, for any word $w$, and conversely, if $wu \slex wv$,
then $u \slex v$.
\item   If $u <_s v$, then $uw <_s v w'$, for any words $w,w'$.
\item \label{cutoff} $u<_s v$ iff there is some $i$ such
that $u[0 \ldots i-1]=v[0 \ldots i-1]$ and $u[0 \ldots i]<_s v[0 \ldots i]$.
\end{enumerate}
\end{prop} 

$\B$ is the two element set $\B=\{0,1\}$
ordered as usual.  The set of words on $\B$,
ordered lexicographically, has the following universal property.
\begin{prop}\label{universal prop}
For any countable linear ordering $(L,<)$ there is a subset $P$ of
$\B^*$ such that  $(L,<)$ is isomorphic to $ (P,\slex)$.
\end{prop} 
{\sl Proof.\/ } 
Any countable linear ordering is isomorphic 
to a subset of the rationals ordered as usual. But
the rationals are isomorphic to the set of words 
on the ordered alphabet $0<1<2$ denoted by the regular 
expression $(0+2)^*1$, since this set
has no first or last element, and between any two words is a third.  But
the ordered set $0<1<2$ is isomorphic to $0 \slex 10 \slex 11$.  Thus,
any countable linear ordering can be embedded in $((0+11)^*10, \slex)$.
 \eop  


A linearly ordered set $(L,\slex)$ is \textit{not} well-ordered if
and only if there is a sequence $(w_n)_{n \geq 0}$ of words in $L$ 
such that $w_{n+1} \slex w_n$, for
all $n$.  In fact, sets of words that are not
 well-ordered by $\slex$ are characterized by the following lemma.
\begin{lemma} If $L \subseteq \zos$ and $(L,\slex)$ is not well-ordered,
then there is an infinite sequence $(w_n)_{n \geq 0}$ of words in $L$
such that
\begin{eqnarray*}
  w_{n+1} &<_s& w_n,
\end{eqnarray*}
for all $n \geq 0$.
\end{lemma}
{\sl Proof.\/ }  Suppose that $(v_n)_{n \geq 0}$ is a countable 
$\slex$-descending chain of words in $L$. 
Then, for each $n$, either $v_{n+1}<_p v_n$ or $v_{n+1}<_s v_n$. Define $u_1=v_1$. 
Since $v_1$ has only finitely many prefixes,
 there is a least integer $k$ such that 
$v_{k+1}<_s v_k<_p ...<_p v_1$. Then 
define $u_2=v_{k+1}<_s u_1$. Similarly, assuming that $u_m$ has been defined 
as $v_{m'}$, for some $m'$, we may define $u_{m+1}$ as the first $v_k$ such that $k>m'$ and $v_k<_s u_m$. \eop

A \textbf{deterministic finite automaton $M$}, DFA for short,
 consists of a finite set $Q$, the ``states'', an element $s \in Q$,
the ``start state'', a finite set $\Sigma $, the ``alphabet'', a function $\delta:Q \x \Sigma \to Q$, the ``transition function'',
and a subset $F$ of $Q$, the ``final states''.  The transition
 function is extended to a function $Q \x \Sigma^* \to Q$
in the standard way:  
\begin{eqnarray*}\label{}
\delta(q, \epsilon ) &:=& q, \quad q \in Q\\
\delta(q, \sigma u) &:=& \delta(\delta( q, \sigma), \ u),
 \quad q \in Q,\ \sigma \in \Sigma,\ u \in \Sigma^*
\end{eqnarray*}
where $\epsilon$ is the empty word.  For $q \in Q,\ u \in 
\Sigma^*$, we write $q.u$ instead of $\delta(q,u)$.
For any state $q$, 
the \textbf{language determined by $q$}, $\L(q)$, is the set
\begin{eqnarray*}\label{}
\L(q) &:=& \{u \in \Sigma^*: q.u \in F\} .
\end{eqnarray*} 
The \textbf{language determined by $M$}, $\L(M)$, is the language determined by
the start state $\L(s)$. 
 We say that a DFA is
\textbf{trim} if for every state $q$, there is some word $u$ such that
$s.u=q$, \textit{and}, there is at most one state $q$ such that $\L(q)= \emptyset  $.
We call a state $q$ such that $\L(q)=\emptyset$ a \textbf{sink state}.

In view of Proposition \ref{universal prop}, from now on
we assume that the alphabet of all DFAs is $\B=\{0,1\}$.

The underlying \textbf{labeled directed graph}, $G(M)$, of a 
DFA $M$ has as vertices the states of $M$;
there is an edge $q \to q'$ labeled $b$ 
if and only if $q.b=q'$, for some $b \in \B$.  A \textbf{strong component} of $M$
is a strong component of $G(M)$.  Recall that two states $q,q'$ are in the same
strong component iff there are paths in $G(M)$ from $q$ to $q'$ and from $q'$ to $q$.
A strong component $c$ is \textbf{nontrivial} if there is at least one edge
$q \to q'$, where both $q,q'$ belong to $c$.  
An edge $q \to q'$ is an \textbf{exit edge} of a strong component $c$
if $q$ belongs to $c$ and $q'$ does not. 
\begin{definition}\label{ord dfa def}
An \textbf{ordinal DFA} is a trim DFA $M$ such that $(\L(M),\slex)$ is well-ordered.
\end{definition} 

\subsection{The characterization theorem}  
\begin{lemma}
\label{inheritance lemma}
Suppose $M$ is an ordinal DFA.
For every state $q$ of $M$,
$(\L(q),\slex)$
is well-ordered.
\end{lemma}
{\sl Proof.\/ }
Suppose that $(w_n)_{n \geq 0}$ is a descending sequence
of words in $(\L(q),\slex)$.  Since $q$ is accessible, there is a word
$v$ such that $s.v=q$. Then $(v w_n)$ is a descending sequence
in $(\L(M),\slex)$, a contradiction. \eop

The next lemma gives a necessary condition that $M$ is an ordinal DFA.
\begin{lemma}[Main Lemma]
  \label{main lemma}
\label{1-lemma}
Let $M$ be an ordinal DFA. For any non sink state $q$, 
if $q$ and $q.0$ are in the same strong component, then $q.1$ is a sink.
\end{lemma}
{\sl Proof.\/ } Suppose, in order to obtain a contradiction,
that $v$ is a word such that $q.1v \in F$.  Let $u$ be a word
such that $(q.0).u=q$.  
For $n \geq 0$, define
\begin{eqnarray*}
  w_n &:=& (0u)^n 1v.
\end{eqnarray*}
Then $w_{n+1} <_s w_n$, for each $n$, and $w_n $ is in $\L(q)$,
contradicting Lemma \ref{inheritance lemma}. 
This contradiction shows $\L(q.1)=\emptyset  $. \eop

In any DFA, a \textbf{recursive state} $q$ is a non sink state
such that
\begin{eqnarray*}
  q.u &=& q,
\end{eqnarray*}
for some nonempty word $u$.

Now we prove the converse to the Main Lemma
\ref{main lemma}.


Suppose that $(\L(M),\slex)$ is not
well-ordered.  Let
\begin{eqnarray*}
\ldots <_s w_{n +1} <_s w_n <_s \ldots <_s w_1
\end{eqnarray*}
be an infinite descending chain in $\L(M)$. 


Say \textbf{position $i$ is active at time $n$} if
\begin{eqnarray*}\label{}
  w_n[0 \ldots i-1] &=& w_{n+1}[0 \ldots i-1]\\
(w_n)_i &=& 1 \\
  (w_{n+1})_i &=& 0.
\end{eqnarray*} 

\begin{remark}
The terminology ``time'' is suggested by the picture
that at the $n$-th click of a clock, the two words
$w_n,w_{n+1}$ are generated, yielding the active position
accounting for the fact that $w_{n+1}<_s w_n$.
\end{remark}
\begin{prop}\label{active bound prop}
There is no upper bound on the active positions.
\end{prop} 
{\sl Proof.\/ } Suppose otherwise. Let $n$ be
a positive integer such that all
active positions are less than $n$.  Then, by part \ref{cutoff} of 
Proposition \ref{elem prop}, there would be an infinite
descending sequence of words of length at most $n$. \eop 

Let $i_0$ be the least position which is active at any time.
\begin{prop}\label{least pos prop}
Position $i_0$ is active at
exactly one time $t_0$.
\end{prop} 
{\sl Proof.\/ } Suppose, in order to obtain a contradiction, 
that $t_0$ is the least time when 
position $i_0$ is active, and that $n>t_0$ is the least time 
after that when
position $i_0$ is active. But then 
$w_{t_0 + 1}[0 \ldots i_0-1]=w_n[0 \ldots i_0 - 1]$ and
$(w_{t_0 + 1})_{i_0}=0$ while $(w_n)_{i_0}=1$, 
showing $w_{t_0+1}<_s w_n$, an impossibility. \eop 

\begin{corollary}\label{}
For all $n>t_0$,
\begin{eqnarray*}\label{}
w_n[0 \ldots i_0] &=& w_{t_0 + 1}[0 \ldots i_0]. \eop 
\end{eqnarray*} 
\end{corollary} 

By considering the descending sequences $(w_n)$, $n > t_0$,
we obtain the following fact.
\begin{prop}\label{}
There is a least position $i_1>i_0$ which is active at a
unique time $t>t_0$. \eop 
\end{prop} 

In fact, the same argument proves the following.
\begin{prop}\label{sequence prop}
There is a unique sequence  $(i_k)_k$ of positions and
a sequence $(t_k)_k$   of times such that for each $k \geq 0$,
\begin{enumerate}
\item $i_0$ is the least position active at any time;
\item $t_0$ is the unique time when $i_0$ is active;
\item $i_{k+1}$ is the least position larger than $i_k$ active at any time larger than $t_k$;
\item $t_{k+1}$ is the unique time larger than $t_{k} $ such that position $i_{k+1}$
is active.
\item 
 For each $k \geq 0$, if $n>t_k$, then
 \begin{eqnarray}\label{extension prop}
 w_{t_k}[0 \ldots i_k-1] &=& w_n[0 \ldots i_k-1].
 \end{eqnarray} 
\end{enumerate} 
\eop 
\end{prop} 

\textbf{Example 1}. 
Consider the sequences
\begin{eqnarray*}\label{}
w_1 &=& 11\\
w_2 &=& 10 \\
w_3 &=& 01  \\
w_4 &=& 00 \\
w_k &=& 00 \ldots , \quad k>4.
\end{eqnarray*} 
Here, $$
i_0 = 0,\ 
t_0 = 2,\ 
i_1 = 1,\ 
t_1 = 3. $$

Then position 0 is active at time 2 and position 1 is active at times 1 and 3.

\textbf{Example 2}. 
For any words $u,v,w$, consider the sequences
\begin{eqnarray*}\label{}
w_1 &=& w1u1v1\\
w_2 &=& w1u1v0 \\
w_3 &=& w1u0v1  \\
w_4 &=& w1u0v0\\
w_5 &=& w0u1v1 \\
w_6 &=& w0u1v0 \\
w_7 &=& w0u0v1 \\
w_k &=& w0u0v1 \ldots , \quad k > 7.
\end{eqnarray*} 
Say $|w|=p,\ |u|=n$ and $|v|=m$.  Then
\begin{eqnarray*}\label{}
i_0 &=& p+1 \\
t_0 &=& 4 \\
i_1 &=& p+1+n + 1 \\
t_1 &=& 6.
\end{eqnarray*} 
From this list of words, we cannot determine $i_2$,
even though 
 position $p+n+m+2$ is active
at times 1, 3, 5.

We are now able to prove the converse of the Main Lemma.
\begin{prop}\label{L(M) not wo prop}
If $(\L(M),\slex)$ is not well-ordered, there is a recursive state $q$ in
the same strong component as $q.0$ and $q.1$ is not a sink.
\end{prop} 
{\sl Proof.\/ } Suppose that $(w_n)_n$ is a descending sequence
in $\L(M)$.  We use the notation of
Proposition \ref{sequence prop}.
Define the state $q_k$ by
\begin{eqnarray*}
  q_k &:=& s.w_{t_k}[0 \ldots i_k - 1],
\end{eqnarray*}
where $s$ is the start state.
 By the pigeonhole principle, there are positive integers $k,p$ with
$q_k = q_{k+p}$.  Then
\begin{eqnarray*}
  w_{t_k} [0 \ldots i_k - 1] &=& w_{t_{k+p}}[0 \ldots i_k - 1]
\end{eqnarray*}
by (\ref{extension prop}, part \ref{cutoff}), so that
\begin{eqnarray*}\label{}
q_k = q_{k+p}&=& q_k.w_{t_{k+p}}[i_k \ldots i_{k+p}-1].
\end{eqnarray*} 
But $(w_{t_{k+p}})_{i_k}=0$, since position $i_k$ is active at
time $t_k$, showing that
$(w_{t_k + 1})_{i_k}=0$ and 
position $i_k$ cannot be active after time $t_k$.
Thus, $q_k$ and $q_k.0$ are in the same strong component.
But $(w_{t_k})_{i_k} = 1$, again, since position $i_k$ is active
at time $t_k$, so that
\begin{eqnarray*}
  q_k.1 &=& s.w_{t_k}[0 \ldots i_k],
\end{eqnarray*}
which is not a sink, since $s.w_{t_k} \in F$.
\eop 
\begin{corollary}\label{not wo cor}
If $M$ is a  trim DFA, then $\L(M),\slex)$ is \textit{not} well-ordered
if and only if there is a recursive state $q$ in
the same strong component as $q.0$ and $q.1$ is not a sink. \eop 
\end{corollary} 
\begin{prop}
\label{algorithm-prop}
Given a trim DFA 
 $M$ with $n$ states,
there is an $O(n^2)$-time algorithm to determine
whether $(\L(M),\slex)$ is well-ordered.
\end{prop}
{\sl Proof.\/ } 
Assume $M$ has $n$ states.
There is a linear time algorithm, say depth-first search, to check, given
states $q,q'$, whether there is a nonempty word $u$ with
$q.u=q'$.
(see e.g., \cite{CLRS}, Chapter 22.)
Then for all states $q$ such that there is a nonempty word
$q.v=q$, check to see that when 
there is a word $u$ with $(q.0).u=q$, 
then $q.1$ is a sink.  This is an $O(n^2)$-time algorithm. 
\eop

\section{Upper bound}\label{ }
In \cite{Heilbrunner} 
it was shown that all nonzero regular well-orderings can be built from
$\bfo$ using the operations of sum and the function $\alpha \mapsto \alpha \x \omega $.
(In \cite{BlCho01}, these operations
on words are axiomatized.)
It follows immediately that the least ordinal which is not regular
is $\omega ^\omega $.  
Another method to obtain
this result uses the equivalence between
regular and automatic ordinals \cite{Delhomme}.  
This note presents another argument, based on the techniques
in \cite{BE}.

\subsection{Ordinals}  
We make some observations on ordinals.
\begin{lemma}
\label{lemma-least}
  The least class $\C$ of ordinals containing 0,1 satisfying
the two conditions
  \begin{itemize}
  \item if $\alpha, \beta \in \C$, then $\alpha + \beta  \in \C$;
\item if $\alpha \in \C$, then $\alpha \x \omega \in \C$
  \end{itemize}
is $ \{\alpha: \alpha< \omega^\omega \} $. \eop 
\end{lemma}

We will use Lemma \ref{lemma-least} to show every ordinal less than 
$\omega^\omega $ is the order-type of $(\L(M),\slex)$, for
some DFA $M$.

\subsection{DFAs and ordinals}  
\begin{definition}
Let $\FA$ be the class of ordinals representable as
the order-type of $(\L(M),\slex) $, for an ordinal DFA $M$.
\end{definition}

We show that $\FA$ has the properties of Lemma
\ref{lemma-least}.
\begin{lemma}
  \begin{itemize}
  \item 0,1 belong to $\FA$.
\item If $\alpha, \beta \in \FA$, then $\alpha+\beta \in \FA$
\item If $\alpha \in \FA$, then $\alpha \x \omega \in \FA$.
  \end{itemize}
\end{lemma}
{\sl Proof.\/ } We prove only the third statement.  Suppose
that $M$ is a DFA with start state $q_1$.  Let $M'$ be the DFA
obtained by adding a new start state $q_0$ to $M$ with the
transitions
\begin{eqnarray*}
  q_0 \cdot 1 &=& q_0 \\
  q_0 \cdot 0 &=& q_1.
\end{eqnarray*}
Otherwise, the states, transitions and final states are those of $M$.
Then the set of words recognized by $M'$ are all those of the
form
\begin{eqnarray*}
  1^n 0 u, \quad u \in \L(M),\ n \geq 0.
\end{eqnarray*}
Thus, if the order-type of $(\L(M),\slex) $ is $\alpha $, the order-type
of $\L(M')$ is
\begin{eqnarray*}
  \alpha + \alpha + \ldots &=& \alpha \x \omega . \eop
\end{eqnarray*}
\begin{cor}
\label{closure-cor}
Every ordinal $\alpha $ less than $\omega^\omega $ is the order-type
of \\
{$(\L(M),\slex)$}, for some ordinal DFA $M$. \eop
\end{cor}

In the remainder of this section
we will prove the converse of Corollary \ref{closure-cor}:
if $\alpha $ is the order-type of $(\L(M),\slex)$,
then $\alpha < \omega^\omega $.


One implication of the Main Lemma \ref{1-lemma} is the following.
\begin{prop}
\label{cycle-prop}
Suppose that $M$ is an ordinal DFA and $q$ is a recursive state. Let
$u_0=u_0^q$ be a shortest nonempty word such that $q.u_0=q$.
Then, if $v$ is any word such that $q.v=q$, then $v$ is some power
of $u_0$, i.e., 
\begin{eqnarray*}
  v &=& u_0^n,
\end{eqnarray*}
for some nonnegative integer $n$.
\end{prop}
{\sl Proof.\/ }
Suppose that $n \geq 0$ is least such that $u_0^{n+1}$ is not a prefix of $v$. Write
\begin{eqnarray*}
  v &=& u_0^n u x w
\end{eqnarray*}
where $u$ is a prefix of $ u_0$, $x \in \B$,  and $ux$ is not a prefix of $u_0$.
If $x=0$, then $u1$ is a prefix of $ u_0$, since $u$ is a proper prefix of $u_0$.
Similarly, if $x=1$, $u0$ is a prefix of $ u_0$.
In either case, $q.u$, $q.u0$ and $q.u1$ 
are in the same strong component, contradicting
the Main Lemma.
 \eop

We will write just $u_0$ rather than $u_0^q$ when the state $q$ is
understood.  

\begin{cor}
\label{behav-cor}
Suppose that $M$ is an ordinal DFA and $q$ is a recursive state in $M$.
Then
$w \in \L(q)$ if and only if
for some $n \geq 0$,
\begin{eqnarray*}
  w &=&  u_0^n  p, 
\end{eqnarray*}
for some prefix $p <_p u_0$
of $u_0$ which belongs to $\L(q)$, or
\begin{eqnarray*}
  w &=&  u_0^n u0v,
\end{eqnarray*}
for some words $u,v$ such that
 $u1 \leq _p u_0$ and $v \in \L(q.u0)$.
\end{cor}
{\sl Proof.\/ } 
It is clear that any word of the above two kinds belongs to $\L(q)$.

Conversely, if the path starting at $q$ determined by the
word $w$ does not leave the loop labeled $u_0$, then
$w = u_0^n p $, for some $n \geq 0$ and some prefix
$p$ of $u_0$ such that $p \in \L(q)$.  Otherwise,
this path leaves the loop after $n$ repetitions
via an exit edge labeled 0, by the Main Lemma.
In this case, $w=u_0^n u0v$, where
$u1 \leq_p u_0$ and $v \in \L(q.u0)$.

This completes the proof.
\eop 

\begin{definition}
\label{PQR def}
Suppose that $M$ is an ordinal DFA and $q$ is a recursive state in $M$.
Define, for each $ n \geq 0$, each 
prefix $u1 \leq _p u_0 = u_0^q$, and each prefix $p$ of $u_0$:
\begin{eqnarray*}
  \P(q, n, p) &:=& \{u_0^n p:\ p \in \L(q)\} \\
  \Q(q, n, u1) &:=& \{u_0^n u0w:\ u0w \in   \L(q)\} \\
  \P(q,n) &:=& \bigcup_{p <_p u_0} \P(q,n,p) \\
  \Q(q, n) &:=& \bigcup_{u1 \leq _p u_0} \Q(q, n, u1) \\
   \R(q,n,u1) &:=& \P(q,n) \cup \Q(q,n,u1)\\
   \R(q,n) &:=&\bigcup_{u1 \leq_p u_0} \R(q,n,u1) .
\end{eqnarray*}
\end{definition}
Note that $\P(q,n), \Q(q,n) $ and $\R(q,n)$ are finite unions.
Also, 
\begin{eqnarray*}\label{}
\R(q,n) &=&  \P(q,n) \cup \Q(q,n).
\end{eqnarray*} 

Thus, by Corollary \ref{behav-cor},
\begin{eqnarray*}
  \L(q) &=& \bigcup_{n \geq 0} \R(q,n).
\end{eqnarray*}

\begin{prop}
\label{hierarchy-prop}
Suppose that $M$ is an ordinal DFA and $q$ is a  
recursive state in $M$.
If $0 \leq n < m$, and if $v \in \R(q,n)$ and $w \in \R(q,m)$, then
$v \slex w$.
\end{prop}
{\sl Proof.\/ } There are several cases.  First, suppose
that $v \in \P(q,n)$. If $w \in \P(q,m)$, then either
\begin{eqnarray*}
  v &=& u_0^n p ,
\end{eqnarray*}
for some $n \geq 0$ and some prefix $p$ of $u_0$ which belongs
to $\L(q)$, and
\begin{eqnarray*}
  w &=& u_0^n u_0^{m-n} p',
\end{eqnarray*}
for some $p' \leq_p u_0$ which belong to $\L(q)$. But then $v<_p w$.

If $w \in \Q(q,m)$, then
\begin{eqnarray*}
  w &=& u_0^n u_0^{m-n} u'0w',
\end{eqnarray*}
so that again $v<_p w$.

Suppose now that $v \in \Q(q,n)$.  If $w \in \P(q,m) \cup \Q(q,m)$,
it is easy to see that $v <_s w$. \eop

\begin{cor}
\label{rec-cor}
Suppose that $M$ is an ordinal DFA and $q$ is a recursive state in $M$. Then
$(\L(q),\slex)$ is  the ordered sum
\begin{eqnarray*}
  (\L(q),\slex) &=& (\R(q,0),\slex)  + \ldots + (\R(q,n),\slex) + \ldots
\end{eqnarray*}
\end{cor}
{\sl Proof.\/ } By
Corollary \ref{behav-cor} and Proposition \ref{hierarchy-prop}. \eop

Let $q $ be a recursive state in an ordinal DFA, and
suppose $u1 \leq_p u_0=u_0^q$.
For a fixed $n \geq 0$, we consider the order-type of 
$\R(q,n, u1) = \P(q,n) \cup \Q(q,n,u1)$.
Note that if $p$ and $u1$ are
prefixes of $u_0$, either $p$ is prefix of $u$ or
$u1$ is a prefix of $p$.

We will find an upper bound for the order-type of $(\R(q,n),\slex)$.
Using the notation of
Definition \ref{PQR def}, for $u1 \leq_p u_0$, define
\begin{eqnarray*}\label{}
A &=& \{u_0^n p: \ p \leq_p u \ \&\ p \in \L(q)\}\\
B &=& \{u_0^n p: \ u1 \leq_p p \leq_p u_0 \ \&\ p \in \L(q)\}.
\end{eqnarray*} 
\begin{prop}
$(\R(q,n,u1),\slex)$ is the ordered sum 
\begin{eqnarray}
\label{R ordered sum}
  (\R(q,n,u1),\slex) &=& (A,\slex) + (\L(q.u_0^nu0),\slex) + (B,\slex),
\end{eqnarray}
so that the order-type of $(\R(q,n,u1),\slex)$ is
\begin{eqnarray}\label{R order type}
k + \alpha + k', 
\end{eqnarray} 
where $k$ is the number of elements in $A$, and
$k'$ is the number of elements in $B$, and
$\alpha $ is the order-type  $(\L(q.u0),\slex)$.
\end{prop}
{\sl Proof.\/ } 
Suppose that $w \in \Q(q,n,u1)$. If
 $v =u_0^np \in A$ then $v <_p w$. Indeed, $w = u_0^n u0 w'$, for some $w' \in \L(q.u0)$.
But since $p \leq_p u$, $v <_p w$.  Similarly, if $v \in B$,
$w <_s v$.  This proves 
(\ref{R ordered sum}).

The order-types of $(\L(q.u0),\slex)$ and $(\L(q.u_0^n u0),\slex)$ are
the same, since 
$q.u_0^n=q$.
We have proved 
(\ref{R order type}).  \eop

Since $\R(q,n)$ is the (non disjoint) union of the sets
$\R(q,n,u1)$, for $u1 \leq_p u_0$, we have the following
result.
\begin{cor}
  \label{ordertype-cor}
For a recursive state $q$, the order-type of $(\R(q,n),\slex)$
is bound\-ed above by a finite sum $\beta_1 + \ldots + \beta_m$, where for
$i=1,\ldots,m$, $\beta_i= k_i + \alpha_i + k'_i$, with
$0 \leq k_i, k'_i < \omega $ and $\alpha_i$ is the order-type
of $(\L(q.u0), \slex)$, for some prefix $u1$ of $u_0$. 
\end{cor}
For later use, we point out the following consequence
of Corollary \ref{ordertype-cor} and Corollary \ref{rec-cor}.
\begin{cor}
  \label{bounded-cor}  
Let $q$ be a recursive state in an ordinal DFA.
Suppose that for each prefix $u1$ of $u_0$,
the order-type of $(\L(q.u0),\slex)$ is less than $\omega^h$, for a positive
integer $h$.  Then the order-type of $\R(q,n)$ is also less than
$\omega^h$, and the order-type of $(\L(q),\slex)$ is at most $\omega^h$.
\end{cor}
{\sl Proof.\/ } The first statement follows from Corollary \ref{ordertype-cor} andthe fact that
ordinals less than $\omega^h$ are closed under
finite sums. The second follows from 
Corollary \ref{rec-cor}.
\eop 

The next definition adopts a similar notion for context-free
grammars from \cite{BE}.
\begin{definition}
Suppose $M$ is any DFA.  For any states $q,q'$, define
  \begin{eqnarray*}
q' \preceq q &\iff & q.v=q',
  \end{eqnarray*}
for some word $v$. Define $[q]=\{q': q \preceq q'\ \&\ q' \preceq q\} $.
\end{definition}
Two states $q,q'$ are \textbf{equivalent} if $q \preceq q$ and
$q'\preceq q$, i.e., they are in the same strong component.
The preorder relation $q\preceq q'$ determines a partial ordering on
the equivalence classes $[q]$:  $[q'] \leq [q]$ if $q' \preceq q$.
\begin{lemma}
Suppose $[q'] \leq [q]$. Then if $M$ is an ordinal DFA,
the order-type of $(\L(q'), \slex)$ is at most that of $(\L(q),\slex)$.
\end{lemma}
{\sl Proof.\/ } Let $v$ be a word such that $q.v=q'$.  Then, for
any word $u \in \L(q')$, the word $vu $ belongs to $ \L(q)$. Thus
\begin{eqnarray*}
  u & \mapsto & vu
\end{eqnarray*}
is an order-preserving map $\L(q') \to \L(q)$. \eop
\begin{definition}
Suppose $M$ is a DFA and $q$ is a state in $M$. The \textbf{height of $q$}
is the number of equivalence classes $[q']$ such that $[q']<[q]$.
\end{definition}
\begin{cor}
\label{proper height cor}
Suppose $M$ is an ordinal DFA.
If $q' \in [q]$, the order-types of $(\L(q),\slex)$ and
$(\L(q'),\slex)$ are the same.   If $q,q'$ have the same height and 
$q' \preceq q$, then $q \preceq q'$.
\end{cor}
{\sl Proof of the last claim.\/ } If there is no path $q' \leadsto q$, then $[q']<[q]$, so that
the height of $q$ is greater than that of $q'$. \eop 
\begin{remark}
In a trim DFA, if there is a sink state, there is a unique one, and
its height is zero.  Conversely, if $q$ is a state of height zero and
$(\L(q),\slex)$ is well-ordered, then $q$ is a sink state. Otherwise, since
both $q.0$ and $q.1$ are in the strong component of $q$, this
contradicts the Main Lemma.
\end{remark}
\begin{theorem}
\label{height-thm}
Suppose that $M$ is an ordinal DFA.
If $q$ is a state of height $h$, then the order-type of $(\L(q),\slex)$
is at most $ \omega^h$.
\end{theorem}
{\sl Proof.\/ } 
 We use induction on $h$.

When $h=0$, $q$ must be a sink state, by the previous remark. Thus,
the order-type of $(\L(q),\slex)$ is 0, and $0<\omega^0=1$.

Assume  $h=1$ and $q$ is not recursive. Then both $q.0$ and
$q.1$ are the sink. If $q \in F$, the order-type
of $(\L(q),\slex)$ is 1; if $q $ is not in $F$, $\L(q)=\emptyset  $,
showing $q$ is a sink, contradicting the assumption
that $M$ is trim.

Assume $h=1$ and $q$ is recursive. Then each exit edge from the strong
component of $q$ labeled either 0 or 1 has the sink
as target.  There must be some final states in the strong component of
$q$, or else $q$ itself is a sink.  Say there are are $k>0$ prefixes of $u_0$ in $F$.
 Since, in this case,
\begin{eqnarray*}\label{}
(\L(q),\slex) &=& (\P(q,0),\slex) + (\P(q,1),\slex) + \ldots 
\end{eqnarray*} 
we see that the order-type of $\L(q),\slex)$ is
\begin{eqnarray*}\label{}
k + k + \ldots &=& \omega .
\end{eqnarray*} 

To complete the induction, assume $h > 1$ and
suppose that if a state has height less than $h$, then the order-type
of its language is at most $\omega^{h'}$, for some
nonnegative integer $h'<h$.  If $q$ has height $h$,
either it is recursive, or not.  If not, the order-type of $q$
is at most $1 + \alpha_0+\alpha_1$, where $\alpha_i$, $i=0,1$, is the order-type of
$(\L(q.i),\slex)$.  Since $q.i$ has height less than $h$, the order-type
of $(\L(q),\slex)$ is at most $1+ \omega^{h-1} \x 2 < \omega^h$.

If state $q$ has height $h$ and $q$ is recursive, then by
Corollary \ref{bounded-cor}, the order-type of $(\L(q),\slex)$ 
is at most
\begin{eqnarray*}
  \omega^{h-1} + \omega^{h-1} + \ldots + \omega^{h-1} + \ldots &=& \omega^{h-1} \x \omega \\
&=& \omega^h. \eop
\end{eqnarray*}

As a consequence of Theorem \ref{height-thm}
and Corollary \ref{closure-cor},
we obtain another proof of the following result.
\begin{cor}
\label{main-cor}
An ordinal $\alpha $ is regular if and only if $\alpha < \omega^\omega $.
\end{cor}
{\sl Proof.\/ } 
By Corollary \ref{closure-cor}, we need prove only that any regular ordinal
is less than $\omega^\omega $.
If $\alpha $ is regular, there is an ordinal DFA $M$ such
that $\alpha $ is the order-type of
$(\L(M),\slex)$.  By Theorem \ref{height-thm}, if $M$ has $n$ states,
the order-type of $(\L(M),\slex)$ is at most $\omega^n$. \eop

\section{Summary}\label{ }

Aside from an alternative proof of the result
in Corollary \ref{main-cor}, we have found a structural
characterization of ordinal DFAs in
Corollary \ref{not wo cor} and an $O(n^2)$-algorithm
to identify them.  It would be interesting to find
a structural characterization of those DFAs
$M$ such that $(\L(M),\slex)$ is 
\begin{itemize}
\item dense, or
\item scattered.
\end{itemize}


\begin{thebibliography}{100}

\bibitem[Ada74]{Adamek}
J.~Adamek.
\newblock Free algebras and automata realizations in the language of
categories.
\newblock \textit{Comment. Math. Univ. Carolinae}, 15(1974), 589--602.


\bibitem[BC01]{BlCho01}
S.L. Bloom and C.Choffrut.
\newblock Long words: the theory of concatenation and $\omega $-power.
\newblock \textit{Theoretical Computer Science}, 259(2001), 533--548.
 
\bibitem[BE10a]{BEa}
S.L. Bloom and Z.\'Esik.
\newblock Algebraic Linear Orderings.
\newblock to appear.

\bibitem[BE10]{BE}
S.L. Bloom and Z.\'Esik.
\newblock Algebraic Ordinals.
\newblock to appear in \textit{Fundamenta Informaticae}.




\bibitem[BE07]{BEbergen}
S.L. Bloom and Z. \'Esik.
\newblock Regular and algebraic words and ordinals. In: \textit{CALCO 2007, Bergen},
\newblock LNCS 4624, Springer, 2007, 1--15.


\bibitem[BE10]{BEmezei}
S.L. Bloom and Z. \'Esik.
A Mezei-Wright theorem for categorical algebras.
\textit{Theoretical Computer Science} 411 (2010) 341--359.
 



\bibitem[CLRS]{CLRS}
T. H. Cormen, C.E. Leiserson, R.L. Rivest, C. Stein.
\newblock \textit{Introduction to Algorithms}, Third Edition.
\newblock The MIT Press. Cambridge, MA., 2009.

\bibitem[Cour78a]{Courcelle78}
B. Courcelle.
\newblock Frontiers of infinite trees.
\newblock \textit{RAIRO Theoretical Informatics and Applications},   12(1978), 319--337.

\bibitem[Cour78]{Courcelle}
B. Courcelle.
\newblock  A representation of trees by languages,
\newblock \textit{Theoretical Computer Science}, 6 (1978), 255--279 and 7(1978), 25--55.


\bibitem[Cour83]{CourcelleFund}
B. Courcelle.
Fundamental properties of infinite trees.
\textit{Theoretical Computer Science},
25(1983), 95--169.


\bibitem[Del04]{Delhomme}
Ch. Delhomm\'e.
\newblock Automaticity of ordinals and of homogeneous graphs.
\newblock C. R. Math. Acad. Sci. Paris  339(2004),  no. 1, 5--10. (in French)




\bibitem[Heil80]{Heilbrunner}
S. Heilbrunner.
\newblock  An algorithm for the solution of fixed-point
equations for infinite words.
\newblock \textit{RAIRO Theoretical Informatics and Applications}, 14(1980), 131--141.


\bibitem[KRS03]{Khoussainovetal}
B. Khoussainov, S. Rubin and F. Stephan.
\newblock On automatic partial orders.
\newblock Proceedings of Eighteenth IEEE Symposium on Logic in Computer Science, LICS, 168-177, 2003.

\bibitem[Roit90]{Roitman}
Judith Roitman.
\newblock \textit{Introduction to Modern Set Theory}.
\newblock Wiley, 1990.

\bibitem[Ros82]{Rosenstein}
J.B. Rosenstein.
\newblock \textit{Linear Orderings}.
\newblock Academic Press, New York, 1982.

\bibitem[Wand79]{Wand}
M.~ Wand.
\newblock Fixed point constructions in order-enriched categories.
\newblock \textit{Theoretical Computer Science}, 8(1979), 13--30.


\end{thebibliography}
\end{document}